\documentclass[epj]{svjour}
\usepackage{graphics}
\begin{document}
%
\title{Theoretical predictions of experimental observables
sensitive to the symmetry energy}
\subtitle{Results of the SMF transport model}
\author{Maria Colonna\inst{1},Virgil Baran\inst{2} \and Massimo Di Toro\inst{1,3}
}                     
%
%
\institute{Laboratori Nazionali del Sud, INFN, via Santa Sofia 62, 
I-95123, Catania, Italy \and
Physics Faculty, University of Bucharest, Romania
\and Physics-Astronomy Dept., University of Catania, Italy} 
\date{Received: date / Revised version: date}
%
\abstract{ In the framework of mean-field based transport approaches, 
we discuss recent results concerning heavy ion 
reactions between charge asymmetric systems, from low
up to intermediate energies. 
We focus on isospin sensitive observables, aiming at extracting information
on the density dependence of the isovector part of the nuclear effective
interaction and of the nuclear symmetry energy. 
For reactions close to the Coulomb barrier, we explore the structure
of collective dipole oscillations, rather sensitive to the low-density
behavior of the symmetry energy. 
In the Fermi energy regime,  
we investigate the interplay between dissipation mechanisms,
fragmentation and isospin  effects. 
At intermediate energies, where regions with higher density and momentum
are reached, we discuss collective flows and their sensitivity to the 
momentum dependence of the isovector interaction channel, which 
determines the splitting 
of neutron and proton effective masses.
Finally, we also discuss the isospin effect 
on the possible phase transition from nucleonic matter to quark 
matter.
Results are critically reviewed, also trying to establish a link, when possible, 
with the outcome of other transport models. 
\PACS{
{25.70.-z}{Low and intermediate energy heavy-ion reactions} \and
{24.60.Ky} {Fluctuation phenomena}     
     } 
} 
\titlerunning{Results of the SMF transport model}
\maketitle
\section{Introduction}

The behavior of nuclear matter in several conditions of density,
temperature and N/Z asymmetry is of fundamental importance for the understanding of 
many phenomena involving nuclear systems and astrophysical compact objects. 
This information can be accessed by mean of heavy ion collision experiments, where transient 
states of nuclear matter spanning a large variety of regimes can be created. 
Actually this study allows one to learn about the corresponding behavior 
of the nuclear effective interaction, which provides the nuclear Equation of State (EOS) 
in the equilibrium limit.  
Over the past years, measurements of experimental observables, like isoscalar collective vibrations, 
collective flows and meson production, have contributed to constrain the EOS of symmetric matter 
for densities up to five time the saturation value \cite{cons}.
More recently, the availability of exotic beams has made it 
possible to explore, in laboratory conditions, new aspects of nuclear structure and dynamics up 
to extreme ratios of neutron (N) to proton (Z) numbers, also opening the way to the investigation of the EOS of
asymmetric matter, which  has 
few experimental constraints. 
Indeed, the isovector part of the nuclear effective interaction and the corresponding
symmetry energy of the EOS (Asy-EOS) are largely unknown as soon as we move away from normal density. 

Nevertheless, this information is essential in the astrophysical context, for the understanding of the 
properties of compact objects such as neutron stars, whose crust behaves as low-density 
asymmetric nuclear matter \cite{Lattimer} and whose core may touch extreme
values of density and asymmetry. 
Moreover, the low-density behavior of the symmetry energy also affects
the structure of exotic nuclei and 
the appearance of new features involving the neutron skin,
whish are currently under intense investigation. \cite{Colo}.

Over the past years,  several observables which are sensitive to the Asy-EOS and testable
experimentally, have been suggested \cite{Isospin01,baranPR,WCI_betty,baoPR08}.
In this article we will review recent results on dissipative collisions in a wide range of beam 
energies, from just above the Coulomb barrier up to the $AGeV$ range,
on the basis of transport theories of the Stochastic Mean Field (SMF) type.  
Low to Fermi energies
 will bring information on the symmetry term around (below) normal density, 
while intermediate energies will probe high density regions.

\section{Transport theories and symmetry energy}

Nuclear reactions are modeled by solving transport equations
based on 
mean field theories, with short range (2p-2h) correlations included via hard nucleon-nucleon
elastic collisions and via stochastic forces, selfconsistently
evaluated from the mean phase-space trajectory, see 
\cite{chomazPR,baranPR}. 
Stochasticity is 
essential in 
order to get distributions as well as to allow for the growth of dynamical 
instabilities. 

In the energy range up to a few hundred $AMeV$, the appropriate tool is the 
so-called Boltzmann-Langevin (BL) equation \cite{chomazPR}:
\begin{equation}
{{df}\over{dt}} = {{\partial f}\over{\partial t}} + \{f,H\} = I_{coll}[f] 
+ \delta I[f],
\end{equation}
 where $f({\bf r},{\bf p},t)$ is the one-body distribution function, 
the semi-classical analog of the Wigner transform of the one-body density matrix, 
$H({\bf r},{\bf p},t)$ the mean field Hamiltonian, 
$I_{coll}$ the two-body collision term 
incorporating the Fermi statistics of the particles,
and 
$\delta I[f]$ its fluctuating part. 
Here we follow the approximate treatment of the BLE introduced in Ref.\cite{Salvo},
the Stochastic Mean Field (SMF) model. 
The numerical procedure to integrate the transport equations is based on the
test-particle method.

Effective interactions (associated with a given EOS) 
can be considered as an input of the transport code
and from the comparison with experimental data one can finally get some hints
on nuclear matter properties. 

We recall that 
the symmetry energy $E_{sym}$ appears in the energy density functional
$\epsilon(\rho,\rho_i) \equiv \epsilon(\rho)+\rho {E_{sym}\over A} (\rho_i/\rho)^2
 + O(\rho_i/\rho)^4 +..$, expressed in terms of total ($\rho=\rho_p+\rho_n$)
and isospin ($\rho_i=\rho_p-\rho_n$) densities.
$E_{sym}$ gets a
kinetic contribution directly from basic Pauli correlations and a potential
part, $C(\rho)$,  from the highly controversial isospin dependence of 
the effective interactions: 
\begin{equation}
\frac{E_{sym}}{A}=\frac{E_{sym}}{A}(kin)+\frac{E_{sym}}{A}(pot)\equiv 
\frac{\epsilon_F}{3} + \frac{C(\rho)}{2\rho_0}\rho 
\end{equation}
($\rho_0$ denotes the saturation density).
The nuclear mean-field, consistently derived from the energy functional, can be written as:
\begin{equation}
U_{q} = A\frac{\rho}{\rho_0}+B(\frac{\rho}{\rho_0})^{\alpha+1} + C(\rho)
\frac{\rho_n-\rho_p}{\rho_0}\tau_q
+\frac{1}{2} \frac{\partial C}{\partial \rho} \frac{(\rho_n-\rho_p)^2}{\rho_0},
\end{equation}
where $\tau_q = +1 (-1)$ for $q=n (p)$.
The isoscalar section is fixed requiring that 
the saturation properties of symmetric nuclear matter, with 
a compressibility modulus around $K=200 MeV$, are reproduced
(which corresponds to the Skyrme SKM* effective interaction).
The corresponding values of the coefficients are
$A=-356.8 MeV$, $B=303.9 MeV$, $\alpha=1/6$.

\begin{figure}
\vskip 0.5cm
\resizebox{0.4\textwidth}{!}{%
  \includegraphics{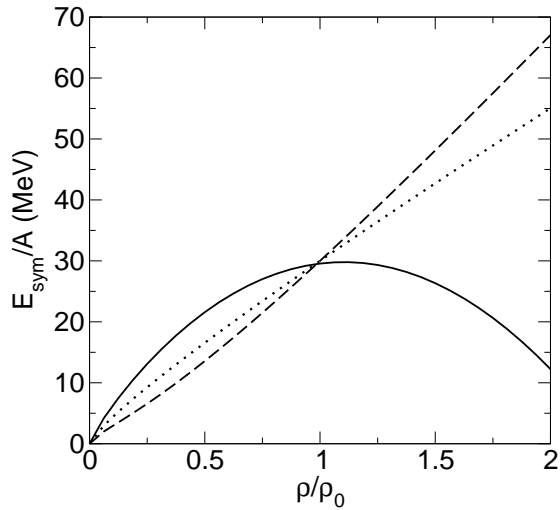}
}
\caption{\label{isoEOS}
Three effective parameterizations of the symmetry 
energy : asystiff (dotted line), asysoft (full line) and
asysuperstiff (dashed line).
}
\end{figure}

The sensitivity of the simulation results is tested against 
different choices of the density dependence of the
isovector part of the effective interaction.
We employ
three different parameterizations of $C(\rho)$: the asysoft,
the asystiff and asysuperstiff respectively, 
see \cite{baranPR} for a detailed description.
The value of the symmetry energy,
$\displaystyle E_{sym}/A $, 
at saturation, as well as
the slope parameter, $\displaystyle L = 3 \rho_0 \frac{d E_{sym}/A}{d \rho} |_{\rho=\rho_0}$,
are reported in Table \ref{table1} (first two columns) for each of these Asy-EOS. Just below the saturation density
the asysoft mean field has a weak variation with density while the asysuperstiff shows
a rapid decrease, see   
figure \ref{isoEOS}.

For protons, the Coulomb interaction is also included in the simulations. 


Surface terms are not explicitly included in the mean-field potential, 
however surface effects are accounted for by considering finite width wave packets
for the test particles employed in the numerical resolution of Eq.(1).
The width is tuned to reproduce the surface energy of nuclei in the ground state
\cite{Alfio}. This method also induces the presence of a surface term
in the symmetry energy. We have checked that properties connected to surface effects, 
such as the neutron skin of neutron-rich nuclei, are in
reasonable agreement with calculations of other
models employing similar interactions \cite{ref17,ref15}.

Momentum-dependent effective interactions may also be implemented into Eq.(1)
and will be considered in the following for the study of observables which are particularly 
sensitive to this ingredient.   In particular, we will discuss results related to the momentum dependence 
of the isovector channel of the interaction, leading
to the splitting of neutron and proton effective masses (see Section 5).



\section{Collective excitations in neutron-rich systems}
One of the important tasks in many-body physics is to understand the emergence of 
collective features as well as their structure in terms of the individual motion of the
constituents.  The experimental characterization and theoretical
description of collective excitations appearing in charge asymmetric and 
exotic systems is a challenge for modern nuclear physics. 

\subsection{New exotic collective excitations}
Recent experiments have provided
several evidences  about the existence of new collective 
excitations in neutron-rich systems, 
but the available information is still
incomplete and their nature is a matter of debate.
In particular, many efforts have been devoted to the study of the 
Pygmy Dipole Resonance (PDR), 
identified as an unusually large concentration of the dipole response at energies
below the values corresponding to the 
Giant Dipole Resonance (GDR). The latter is one of the most prominent and robust collective
motions, present in all nuclei, whose centroid position varies, for
medium-heavy nuclei, as $80 A^{-1/3} MeV$.
 From a comparison of the
available data for stable and unstable $Sn$ isotopes a correlation
between the fraction of pygmy strength and isospin asymmetry was
noted \cite{ref2}. In general the exhausted sum-rule
increases with the proton-to-neutron asymmetry. This behavior was
related to the symmetry energy properties below saturation and
therefore connected to the size of the neutron skin
\cite{ref3,ref4,Colo}.

\begin{table*}
\begin{tabular}{|l|r|r|r|r|r|} \hline
asy-EoS       & $E_{sym}/A$    & L(MeV)  & $R_n$(fm) & $R_p$(fm) & $\Delta R_{np}(fm)$  \\ \hline
asysoft       &     29.9                 & 14.4    & 4.90  & 4.65    & 0.25 \\ \hline
asystiff      &     28.3                 & 72.6    & 4.95  &  4.65   & 0.30 \\ \hline
asysupstiff   &     28.3                 & 96.6    & 4.96  &  4.65   & 0.31 \\ \hline
\end{tabular}
\caption{The symmetry energy at saturation (in $MeV$), the slope parameters, neutron rms radius,
protons rms radius, neutron skin thickness of $^{132}$Sn for the three Asy-EOS.}
\label{table1}
\end{table*}

 In spite of the theoretical progress in the interpretation of
this mode and new experimental
information \cite{ref6,ref7,ref8,ref9}, a number of critical
questions concerning the nature of the PDR still remains.
Here we want to address the important issue related
to the collective nature of the PDR in connection with the role of the symmetry energy.

A microscopic, self-consistent study of the collective features
and of the role of the nuclear effective interaction
upon the PDR can be performed within the Landau theory of Fermi liquids.
This is based on two coupled Vlasov kinetic equations (see Eq.(1), neglecting
two-body correlations) for neutron and proton one-body
distribution functions $f_q(\vec{r},\vec{p},t)$ with $q=n,p$, 
and was applied  quite successfully in describing various features of the GDR,
including pre-equilibrium dipole excitation in fusion reactions 
\cite{ref11}, see Subsection 3.4.
However, it should be noticed that within such a semi-classical description
shell effects are absent, certainly important in shaping the fine structure
of the dipole response \cite{ref12}.
By solving numerically the Vlasov equation, in the absence of Coulomb interaction, 
Urban \cite{ref13} evidenced from the study of the total dipole moment $D$ a collective response 
around $8.6$ $MeV$ which was identified as a pygmy mode. It was pointed out, from the properties of
transition densities and velocities, that the PDR can be related to one of the low-lying modes 
associated with isoscalar toroidal excitations, providing  indications about its isoscalar character.
Here, considering in the transport simulations also the Coulomb interaction, we
can investigate in a complementary way the collective nature of the PDR by studying the 
dynamics of the pygmy degree of freedom, that is usually associated with the neutron excess in the nuclear surface \cite{ref17}. 
Moreover, we can explore the
isoscalar character of the
mode by a comparative analysis employing three different density parametrizations of the 
symmetry energy.

\subsection{Ingredients of the simulations}
We consider the
neutron rich nucleus $^{132}Sn$ and we
determine its ground state configuration as the equilibrium (static)
solution of Eq.(1). 
Then  proton and neutron densities
$\displaystyle \rho_q(\vec{r},t)=\int \frac{2 d^3 {\bf p}}{(2\pi\hbar)^3}f_q(\vec{r},\vec{p},t)$
can be evaluated.
As an additional check of our initialization procedure,
the neutron and proton mean square radii
\begin{equation}
\langle r_q^2 \rangle = \frac{1}{N_q} \int r^2 \rho_q(\vec{r},t) d^3 {\bf r},
\end{equation}
as well as the skin thickness
$\displaystyle \Delta R_{np}= \sqrt{\langle r_n^2 \rangle}-\sqrt{\langle r_p^2 \rangle}$,
were also calculated in the ground state and shown in Table \ref{table1}.

The values obtained with the semi-classical approach
are in a reasonable agreement with those reported by employing other
models for similar interactions \cite{ref15}.
The neutron skin thickness is increasing with the slope parameter,
as expected from a faster reduction of the symmetry term on the surface 
\cite{ref3,baranPR}.
This feature has been discussed in detail in \cite{Colo}.

To mimic the excitation induced by nuclear reactions, we introduce an initial perturbation
in our system.  
  To inquire on the collective properties of the pygmy dipole, 
we first boost
along the $z$
direction all excess neutrons ($N_e = 32$) and, in opposite direction, all core
nucleons, while keeping the {\rm c.m.} of the nucleus at rest (Pygmy-like
initial conditions).
The excess neutrons were identified as the most
distant $N_e=32$ neutrons from the nucleus c.m.. Then the system is left to
evolve and the evolution of the collective coordinates $Y$, $X_c$ and $X$,
associated with the  different isovector dipole modes (pygmy, core and total dipole)
is followed for $600 fm/c$
by solving numerically the equations (1).
The total dipole moment $D$ is linked to the dipole $D_y$ and the core dipole $D_c$ 
by the following relation:  
\begin{equation}
\vec{D}=\frac{N Z}{A} \vec{X}= \frac{Z~ N_c}{A_c} \vec{X}_{c} + \frac{Z~ N_e}{A} \vec{Y} \equiv 
\vec{D}_c+\vec{D}_y,
\label{diptot}
\end{equation}
where $N_c$ and $A_c$ denote neutron and mass number of the core, respectively.


\subsection{Results for dipole oscillations}
\begin{figure}
\hskip 1.cm 
\resizebox{0.4\textwidth}{!}{%
  \includegraphics{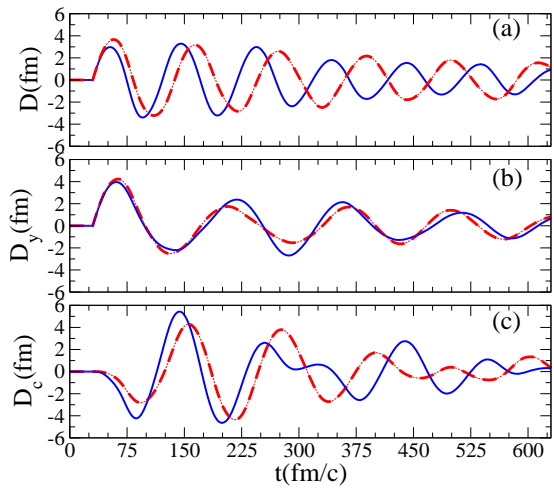}
}
\caption{(Color online) The time evolution of the total dipole $D$ (a),
of the dipole $D_y$ (b) and of core dipole $D_c$ (c),
for asysoft (the blue (solid) lines) and asysuperstiff (the red (dashed) lines) EOS. 
Pygmy-like initial excitation \cite{ref17}.}
\label{diptime}
\end{figure}

In figure  \ref{diptime} we plot the time evolution of 
$D_y$, $D$ and  $D_c$,
for two Asy-EoS.
Apart from the quite undamped oscillations of the $Y$ coordinate, 
we also remark that the core does not remain inert.
Indeed, while $D_y$ approaches its maximum
value, an oscillatory motion of the dipole $D_c$ initiates and
this response is symmetry energy dependent: the larger is the slope
parameter $L$, the more delayed is the isovector core reaction.
 This can be explained in terms of low-density (surface)
contributions to the vibration and therefore of the density behavior
of the symmetry energy below normal density: a larger L corresponds
to a larger neutron presence in the surface (see Table I) and so to a smaller
coupling to the core protons.
We see that the total dipole $D(t)$ is strongly affected by the presence of
isovector core oscillations,
mostly related to the isovector part of the effective interaction.
Indeed, $D(t)$ gets a higher oscillation frequency with respect to
$D_y$, sensitive to the Asy-EOS. The fastest
vibrations are observed in the asysoft case, which gives the largest
value of the symmetry energy below saturation. On the other hand, the
frequency of the pygmy mode seems to be not much affected by the
trend of the symmetry energy below saturation, see also next figure
\ref{dipspectrum}, clearly showing the different nature,
isoscalar-like, of this oscillation. For each Asy-EOS we calculate the
power spectrum of $D_y$:
$\displaystyle |D_y (\omega)| ^2 = |\int_{t_0}^{t_{max}} D_y(t) e^{-i\omega t} dt|^2$,
and similarly for $D$. The results are shown in figure
\ref{dipspectrum}. The position of the centroid corresponding to
the GDR shifts toward larger values when we move from asysuperstiff (largest slope parameter $L$)
to asysoft EOS.
As it clearly appears from Fig.3 (bottom), 
the energy centroid associated with the PDR is situated below the GDR peak, at around $8.5 MeV$,
quite insensitive to the Asy-EOS,
pointing to an isoscalar-like nature of this mode.
 Hence the structure of the
dipole response can be explained in terms of the development of
isoscalar-like (PDR) and isovector-like (GDR) modes 
\cite{ref16}.
We observe that the GDR energy centroid is underestimated in comparison with experimental data, a fact 
probably related to the choice
of the interaction, which has not an effective mass 
\cite{surNPA1988}. On the other hand, the PDR energy centroid looks in better agreement with experimental 
observations \cite{Virgil_new}. 
This may suggest that the PDR peak energy is less
affected by the momentum dependence of the effective interaction,
however  more complete analyses should be performed.
\begin{figure}
\hskip 1.cm
\resizebox{0.4\textwidth}{!}{%
  \includegraphics{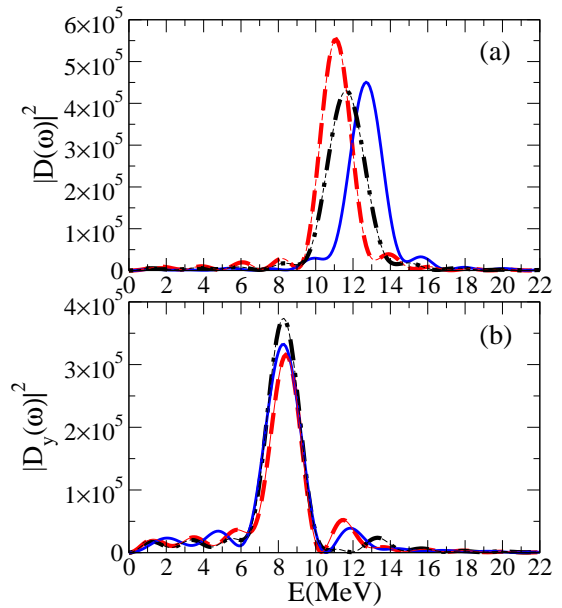}
}
\caption{\label{dipspectrum}(Color online) The power spectrum of total dipole (a) and
of  the dipole $D_y$ (b) (in $fm^4/c^2$), for asysoft
(the blue (solid) lines), asystiff (the black (dot-dashed) lines) and asysuperstiff (the red (dashed) lines)
EOS. Pygmy-like initial conditions \cite{ref17}.
}
\end{figure}

Both modes are excited in the considered pygmy-like initial conditions. Looking
at the total dipole mode direction, that is close to the isovector-like normal mode,
one observes a quite large contribution in the GDR region. 
On the other hand, 
although the pygmy mode has a more complicated structure \cite{ref13},  
the $Y$ direction appears  closely related to it. 
Indeed, looking at $D_y$, a larger
response amplitude is detected in the pygmy region, see figure 3 (bottom).

\begin{figure}
\hskip 1.cm
\resizebox{0.4\textwidth}{!}{%
  \includegraphics{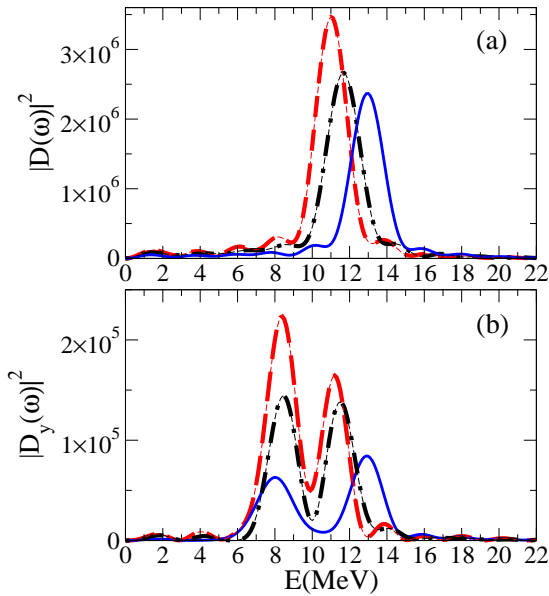}
}
\caption{\label{gdrspectrum}(Color online) The same as in figure \ref{dipspectrum} but for a GDR-like
initial excitation \cite{ref17}.}

\end{figure}
The results crucially depend on the initial excitation of the system. 
Let us consider the case of a  GDR-like excitation, corresponding
to  a boost of all neutrons against all protons, keeping the  {\rm c.m.} at rest.
Now the initial conditions favor the isovector-like mode and
even in the $Y$ direction we observe a sizable contribution in the
GDR region, see the Fourier spectrum of $D_y$ in figure \ref{gdrspectrum}. 
Hence a part of the $N_e$ excess neutrons is involved in a GDR type motion,
and the relative weight depends on the symmetry
energy: more neutrons 
are involved in the pygmy mode in the 
asysuperstiff EOS case, in connection to the larger slope $L$.
We notice that a larger slope $L$ implies a larger coupling between isoscalar
and isovector oscillations in asymmetric matter. As a consequence, 
even for a GDR-like excitation, 
the response in the PDR region will be of greater amplitude for a larger L. This
is also observed in the Fourier spectrum of the total dipole mode
$D$ (figure 4, top). 
We notice that the strength function is related 
to $Im(D(\omega))$ \cite{ref18} and then the corresponding
cross section can be calculated. 
Our estimate of the integrated cross section 
over the PDR region represents  $2.7 \%$
for asysoft, $4.4 \%$ for asystiff and $4.5 \%$ for asysuperstiff,
 out of the total cross section. 
We also remark that a larger slope $L$ is associated with a 
larger neutron skin, thus building a correlation between the
energy weighted sum rule (EWSR) exhausted by the PDR and the neutron skin 
thickness \cite{Virgil_new},
in agreement with the results of \cite{ref19}.

\subsection{The prompt dipole $\gamma$-ray emission in dissipative collisions}
The low-density behavior of the symmetry energy can also be explored 
looking at pre-equilibrium dipole excitations
in dissipative charge asymmetric reactions around 10 $AMeV$.  
The possibility of an entrance channel bremss trahlung dipole radiation
due to an initial different N/Z distribution was suggested at the beginning
of the nineties \cite{ChomazNPA563}. 
After several experimental evidences, in fusion as well as in deep-inelastic
reactions, \cite{PierrouPRC71,medea,Petro} (and references therein),  
the process is 
now well understood and stimulating new perspectives
are coming from the use of radioactive beams.

During the charge equilibration process taking place
 in the first stage of dissipative reactions between colliding ions with
 different N/Z
ratios, a large amplitude dipole collective motion develops in the composite
dinuclear system, the so-called Dynamical Dipole mode. This collective dipole
gives rise to a prompt $\gamma $-ray emission which depends
on the absolute value of the initial amplitude, $D(t=0)$,
on the fusion/deep-inelastic dynamics and 
on the symmetry term, below saturation, that is acting as a restoring
force. Indeed this oscillation develops in the low density interface between
the two colliding ions (neck region). 

\begin{figure}
\hskip 0.5cm
\resizebox{0.5\textwidth}{!}{%
\includegraphics{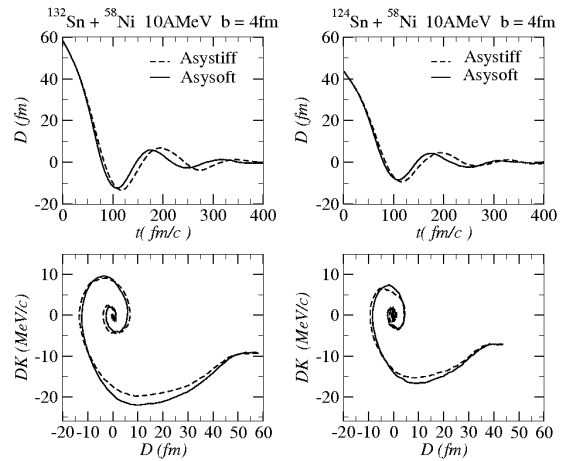}} 
\caption{Dipole Dynamics at 10 $AMeV$, $b=4fm$ centrality. 
Left panels: exotic ``132'' system. Upper panel: time evolution of dipole moment 
D(t) in coordinate
space; Lower panel: dipole phase-space correlation (see text).
Right panels: same as before for the stable ``124'' system.
Solid lines correspond to asysoft EOS, the dashed to asystiff EOS \cite{ref11}.}
\label{dip}
\end{figure}

A detailed description is obtained in the mean field transport approach
\cite{BaranPRL87}.
One can follow the time evolution
of the dipole moment
in the $r$-space,
 $D(t)= \frac{NZ}{A} ({R_{Z}}- {R_{N}})$ and in
$p-$space, $DK(t)=(\frac{P_{p}}{Z}-\frac{P_{n}}{N})$, 
being $R_p$, $P_{p}$
($R_n$, $P_{n}$) the centers of mass in coordinate and momentum space for protons (neutrons).
A nice "spiral-correlation"
clearly denotes the collective nature
 of the mode, see figure \ref{dip} for $Sn + Ni$ reactions at 10 MeV/A.

The ``prompt'' photon emission probability, with energy 
$E_{\gamma}= \hbar \omega$,
 can be estimated applying
a bremsstrahlung approach
 to the dipole evolution given from the BL equation (1):
\begin{equation}
\frac{dP}{dE_{\gamma}}= \frac{2 e^2}{3\pi \hbar c^3 E_{\gamma}}
 |D''(\omega)|^{2}  \label{brems},
\end{equation}
where $D''(\omega)$ is the Fourier transform of the dipole acceleration
$D''(t)$. We remark that in this way it is possible
to evaluate, in {\it absolute} values, the corresponding pre-equilibrium
photon emission.

We must add a
couple of comments of interest for the experimental selection of the Dynamical
Dipole: i) The centroid is always shifted to lower energies (large
deformation of the dinucleus); ii) A clear angular anisotropy should be present
since the prompt mode has a definite axis of oscillation
(on the reaction plane) at variance with the statistical $GDR$.
These features have been observed in recent experiments \cite{medea}.



The use of unstable neutron rich projectiles would largely increase the
effect, due to the possibility of larger entrance channel asymmetries.
Indeed one can  notice in figure \ref{dip} the large amplitude of the first oscillation for the ``132'' system. 
We also remark the delayed dynamics for the asystiff EOS related to a weaker 
isovector restoring force. 
\begin{figure}
\hskip 0.5cm
\resizebox{0.4\textwidth}{!}{%
\includegraphics{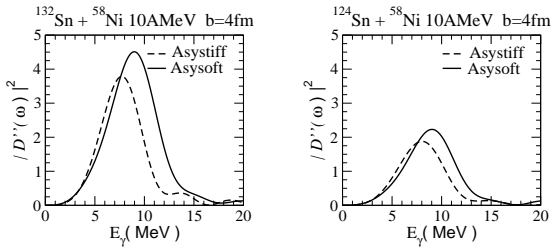}} 
\caption{Left panel: exotic ``132'' system. Power spectra of the 
dipole acceleration at  $b=4$fm (in $c^2$ units).
Right panel: corresponding results for the stable ``124'' system.
Solid lines correspond to asysoft EOS, the dashed to asystiff EOS \cite{ref11}.}
\label{yield1}
\end{figure}
In figure \ref{yield1} (left panel) we report the power spectrum, 
$\mid D''(\omega) \mid^2$, in semicentral
``132'' reactions, for different Asy-EOS choices.
The gamma multiplicity is simply related to it, see Eq.(\ref{brems}).
The corresponding results for the stable ``124''  system are drawn
in the right panel.
As expected from the larger initial charge asymmetry, we clearly see an 
increase 
of the prompt dipole emission for the exotic
n-rich beam. Such entrance channel effect allows also for 
a better observation of the Asy-EOS dependence. 
The asystiff case corresponds to a  
lower value of the centroid, $\omega_0$,  and to a reduced total yield, as shown 
in figure \ref{yield1}. 
In fact, 
in the asystiff case we have a weaker restoring force 
for the dynamical dipole
in the dilute ``neck'' region, where the symmetry energy is smaller 
\cite{dipang08}, see figure 1. 
The sensitivity of $\omega_0$ to the stiffness of the symmetry
energy is more easily identified increasing $D(t_0)$, i.e. using exotic,
more asymmetric beams.
Moreover, a good  sensitivity to the symmetry energy is also observed for dipole
oscillations in more peripheral collisions \cite{carmelo_new}, where low-density surface
contributions become more important. 

The pre-equilibrium dipole radiation angular distribution is the result of the 
interplay between the dinuclear rotation and the collective oscillation life-time. 
Since the latter is sensitive to the Asy-EOS, with a prompter dipole emission in the
soft case, one also expects a sensitivity to the Asy-EOS of the
anisotropy, in particular for high spin event selections \cite{dipang08}.



\section{Isospin equilibration and fragmentation mechanisms at Fermi energies}

We now move to discuss reaction mechanisms occurring at the so-called Fermi energies (30-60 $AMeV$),
where two-body correlations and fluctuations start to play an important role.
In this energy range, reactions between charge asymmetric systems are charecterized by  
a direct isospin transport in binary events (isospin diffusion).
This process also involves the low density neck region and is sensitive to
the low density behavior of $E_{sym}$, see Refs.\cite{tsang92,isotr07} and references therein.
Moreover, it is now quite well established that the largest part of the reaction
cross section for dissipative collisions at Fermi energies goes
through the {\it Neck Fragmentation} channel, with intermediate mass
fragments (IMF) directly
produced in the interacting zone in semiperipheral collisions on short
time scales \cite{wcineck}. It is possible to predict interesting 
isospin transport effects also  for this 
fragmentation mechanism. Clusters are still formed in a dilute
asymmetric matter but always in contact with the regions of the
projectile-like and target-like remnants almost at normal densities,
thus favoring the neutron enrichment of the neck region (isospin migration).   

Results on these mechanisms, obtained with the
SMF model, are discussed below. 
However, the main role  of the isospin degree of freedom on the dynamics of a nuclear reaction
can be easily understood, within the hydrodynamical limit,  considering the behavior
of neutron and proton 
chemical potentials as a function of density $\rho$ and 
asymmetry 
$\beta = (N-Z)/A$ \cite{isotr05}. The $proton/neutron$ currents can be expressed as
\begin{equation}
{\bf j}_{p/n} = D^{\rho}_{p/n}{\bf \nabla} \rho - D^{\beta}_{p/n}{\bf \nabla} \beta ,
\end{equation}
with $D^{\rho}_{p/n}$ the drift, and
$D^{\beta}_{p/n}$ the diffusion coefficients for transport, which are given 
explicitely
 in Ref. \cite{isotr05}. Of interest for the study of isospin effects 
are the differences of currents 
between protons 
and neutrons which have a simple relation to the density dependence of the 
symmetry energy
\begin{eqnarray}
D^{\rho}_{n} - D^{\rho}_{p}  & \propto & 4 \beta \frac{\partial E_{sym}}
{\partial \rho} \,
 ,  \nonumber\\
D^{\beta}_{n} - D^{\beta}_{p} & \propto & 4 \rho E_{sym} \, .
\label{trcoeff}
\end{eqnarray}
Thus the isospin transport due to density gradients (isospin migration) 
depends on the slope of the symmetry energy, or the symmetry pressure, 
while the 
transport due to isospin concentration gradients (isospin diffusion) 
depends on
 the absolute value of the symmetry energy. 
Hence transport phenomena in nuclear reactions appear directly linked to the 
EOS properties. 



\subsection{The isospin transport ratio }
In peripheral and semi-peripheral reactions, 
 it is interesting to look at the asymmetries of the various parts 
of the interacting system in the exit channel:
emitted particles,  projectile-like (PLF) 
and target-like fragments (TLF), and in  the  case of ternary (or higher
multiplicity) events,  
IMF's.
 In particular, one can  study  the
 so-called isospin transport ratio, which is defined as
\begin{equation}
R^x_{P,T} = \frac{2(x^M-x^{eq})}{(x^H-x^L)}~,
\label{imb_rat}
\end{equation}
with $x^{eq}=\frac{1}{2}(x^H+x^L)$.
 Here, $x$ is an isospin sensitive quantity
that has to be investigated with respect to
equilibration.   We consider primarily the asymmetry 
$\beta = (N-Z)/A$,
but also other quantities, such as isoscaling coefficients, ratios of 
production of light
 fragments, etc, can be of interest \cite{WCI_betty,Colonna_IMF,Colonna_PRL}. 
The indices $H$ and $L$ refer to symmetric reactions
between a
heavy  ($n$-rich) and a light ($n$-poor)  system, while $M$ refers to the
mixed reaction.
$P,T$ denote the rapidity region, in which this quantity is measured, in
particular the
PLF and TLF rapidity regions. Clearly, this ratio is $\pm1$ in
the projectile
and target regions, respectively, for complete transparency, and oppositely
for complete
rebound, while it is zero for complete equilibration.
\begin{figure}[h]
\resizebox{0.5\textwidth}{!}{%
\includegraphics{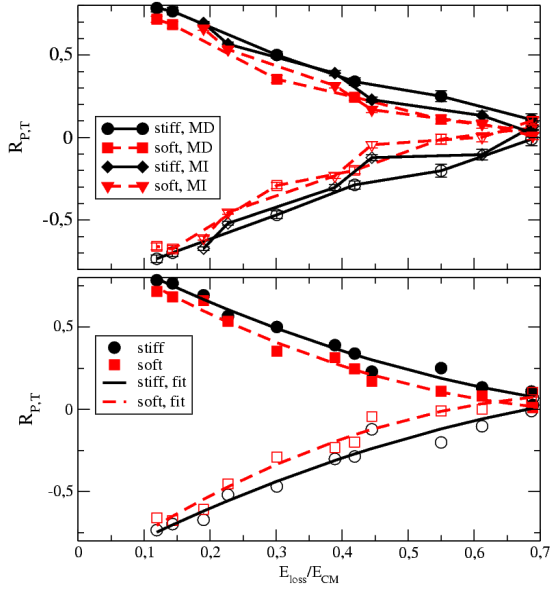}} 
\caption{\label{imb_eloss}
Isospin transport ratios as a function of relative energy loss. 
Upper panel: separately for 
stiff (solid) and soft (dashed) Asy-EOS, and for 
two parameterizations of the isoscalar part of the interaction: 
MD 
(circles and squares) and MI 
(diamonds and triangles), in the projectile region (full symbols)
 and the target region 
(open symbols).
Lower panel: quadratic fit to all points for the stiff (solid), resp.
 soft (dashed) 
Asy-EOS \cite{isotr07}.}
\end{figure}  


In a simple model one can show that the isospin transport ratio mainly depends on two
quantities: the strength of the symmetry energy and the interaction
time between the two reaction partners.
Let us take, for instance, the asymmetry $\beta$ of the PLF (or TLF) as the
quantity $x$.
At a first order approximation, in the mixed reaction this quantity relaxes
towards
its complete equilibration value, $\beta_{eq} = (\beta_H + \beta_L)/2$, as
\begin{equation}
\label{dif_new}
\beta^M_{P,T} = \beta^{eq} + (\beta^{H,L} -  \beta^{eq})~e^{-t/\tau},
\end{equation}
where $t$ is the time elapsed while the reaction partners stay in contact
(interaction time) and the damping $\tau$ is mainly connected to the strength 
of the symmetry energy \cite{isotr07}. 
Inserting this expression into Eq.(\ref{imb_rat}), one obtains
$ R^{\beta}_{P,T} = \pm e^{-t/\tau}$ for the PLF and TLF regions, respectively.
Hence the isospin transport ratio can be considered as a good observable to 
trace back the strength
of the symmetry energy from the reaction dynamics
provided a suitable selection of the interaction time is performed.
The centrality dependence of the isospin ratio, for Sn + Sn collisions at 35 and 50 $AMeV$,
has been investigated in experiments as well as in theory
\cite{tsang92,bao_prl2005,isotr07,yingxun}.
Information about the stiffness of the symmetry energy, pointing to a $L$ value in
the region between 40 and 80 Mev (for a symmetry energy coefficient around 30 MeV at saturation)
has been extracted from
the analysis presented in \cite{yingxun}, based on other transport models.  

Here we investigate more in detail the relation between charge equilibration, 
interaction times and thermal equilibrium.  
Longer interaction times should be correlated to
a larger 
dissipation and kinetic energy loss.  It is then natural to look at the correlation between
the isospin transport ratio and the total kinetic energy loss.
In this way one can also better isolate the role of the symmetry energy in determining $R$
from dynamical effects 
connected to the overall reaction dynamics (isoscalar effects),   see \cite{isotr07}.
In the calculations, we will employ two parametrizations of the isoscalar part of the
nuclear interaction (one without momentum dependence, MI, see Eq.(3) and
one with Momentum Dependence, MD, see \cite{GalePRC41,isotr07}) and two symmetry energy parametrizations.

It is seen in figure \ref{imb_eloss} (top) that the curves for the 
asysoft EOS (dashed) are 
generally lower in the projectile region
 (and oppositely for the target region), i.e. show 
more equilibration, than those for the asystiff EOS, due to the higher value
of the symmetry energy at low density. 
To emphasize 
this trend, all  the values for the
stiff (circles) and 
the soft (squares) Asy-EOS, corresponding to different impact
parameters, beam energies and also to the two possible parametrizations of the isoscalar part of the
nuclear interaction (MD and MI), are collected together in the bottom part of the figure. 
Though MD interactions lead to faster dynamics and shorter interaction times, 
one can see that, using the energy loss as a measure 
of the interaction time,  all the points essentially follow a given line,
depending only on the symmetry energy parameterization adopted.  
It should be noticed that in the MD calculations shown here 
the momentum dependence of the isovector channel of the effective
interaction, leading to the splitting of neutron and proton effective
masses, has been neglected. However, we have checked that the latter would modify
the isospin transport ratio only by a few percent.

It is observed,
 that there is a systematic effect of the symmetry energy of the order 
of about 20 percent, 
which should be measurable. 
Moreover, we notice that the quantity $R$ is a rapidly decreasing function of the degree
of dissipation, $E_{loss}$,  reached in the collision. This can be explained 
in terms of 
dissipation mechanisms maily due to mean-field effects, as predicted by the
SMF model. Indeed, according to a mean-field picture, a significant degree of
thermal equilibrium 
(i.e. a considerable $E_{loss}$) implies a rather long contact time between
the two reaction partners, thus certainly leading to isospin equilibration,
which needs a short time scale to be reached.   
The correlation suggested in figure \ref{imb_eloss}
should represent 
a general feature of isospin diffusion, expected on the basis of  
mean-field dominated mechanisms.
On the other hand, when other dissipation mechanisms (like a sizable particle and fragment emission)
become important, the energy loss is not directly related to the interaction
time between PLF and TLF and isospin equilibration may not occur even in very
dissipative events.  
Thus it would be of great 
interest to verify  experimentally the kind of correlations presented in figure 7.
A similar analysis, exploiting the N/Z content of the light particle
emission to extract the PLF asymmetry, has been performed in \cite{Emmanuelle},
pointing to a stiff behavior (L$\approx$75 MeV) of the symmetry energy. 


\subsection{Isospin dynamics in neck fragmentation at Fermi energies}

\begin{figure}[h]
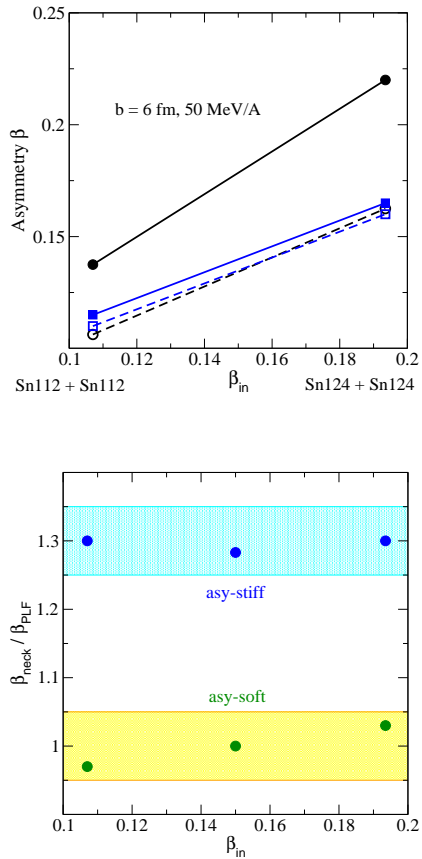

\vskip 0.5cm
\hskip 1.5cm
\resizebox{0.3\textwidth}{!}{%
\includegraphics{fig03_a.eps}} 
\vskip 1.0cm
\hskip 1.5cm
\resizebox{0.3\textwidth}{!}{%
\includegraphics{fig03_b.eps}} 
\caption{
Top panel: asymmetry of IMF's (circles) and PLF-TLF (squares), as a function of the
system initial asymmetry, for two Asy-EOS choices: asystiff (full lines) and
asysoft (dashed lines).  
Bottom panel: 
Ratio between the neck IMF and the PLF asymmetries, as a function of the system
initial asymmetry. The bands indicate the uncertainty in the calculations \cite{Zhang}.}
\label{nzphi}
\end{figure}

In presence of density gradients, as the ones occurring 
when a low-density neck region is formed between the two reaction
partners in semi-peripheral collisions,  the isospin transport
is mainly ruled by the density derivative of the symmetry energy  and so
we expect a larger neutron flow to
 the neck clusters for a stiffer symmetry energy around saturation  
\cite{baranPR}. This mechanism leads to the neutron enrichment of the
neck region (isospin migration). This effect is
shown in figure \ref{nzphi}, where the asymmetry of the neck and
PLF-TLF regions, obtained in neutron-rich reactions at 50 $AMeV$, 
are plotted for two Asy-EOS choices. 

From the experimental point of view, 
a new analysis has been recently published on Sn+Ni data at $35~AMeV$
by the Chimera Collab.\cite{defilposter}.
A strong correlation between neutron enrichement and fragment alignement (when the 
short emission time selection is enforced) is seen, that points to 
a stiff behavior of the symmetry energy ($L\approx 75~ MeV$), 
for which a large neutron
enrichment of neck fragments is seen (figure 8, top). 


In order to build observables less affected by secondary decay effects, 
in figure 8 (bottom)  we consider the ratio of the asymmetries of the IMF's to those of the
residues ($\beta_{res}$) for stiff and soft Asy-EOS, as given by SMF results. 
This quantity 
can be roughly estimated
on the basis of simple energy balance considerations.
By imposing to get a  maximum  (negative) variation of 
$E_{sym}$ when transfering the neutron richness from
PLF and TLF towards the neck region, one obtains \cite{isotr07}: 
\begin{equation}
\frac{\beta_{IMF}}{\beta_{res}} 
= \frac{E_{sym}(\rho_R)}{E_{sym}(\rho_I)} 
\end{equation}
From this simple argument the ratio between the IMF and residue asymmetries should
depend only on symmetry energy properties and, in particular, on the different  
symmetry energy values corresponding to the residue and neck densities ($\rho_R$ and
$\rho_I$),
as appropriate 
for isospin migration.  
It should also be larger than one, more so for the asystiff than
for the asysoft EOS.
It is seen indeed in figure 8 (bottom part), that this ratio 
is nicely dependent on the Asy-EOS only (being larger in the asystiff case) 
and not on the system considered.
If final asymmetries were affected in the same way by secondary evaporation 
in the case of neck and PLF fragments, then one could directly compare the
results of figure 8 (bottom) to data.  However, due to the different size and
temperature of the neck region with respect to PLF or TLF sources, 
de-excitation effects should be carefully checked with the help of
suitable decay codes.

\begin{figure}
\vskip 2.cm
\resizebox{0.5\textwidth}{!}{%
  \includegraphics{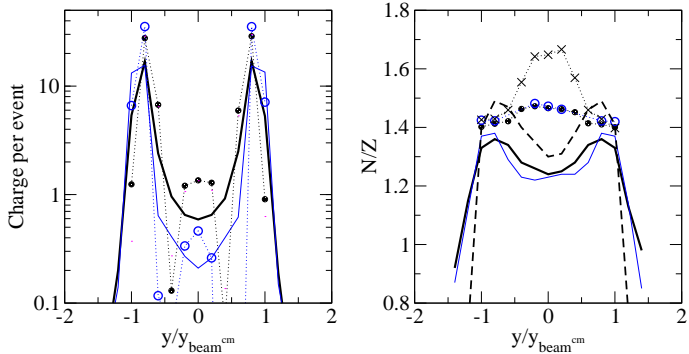}
}
\caption{\label{confro}
Left panel: Average total charge per event, associated with IMF's, as a function of the reduced 
rapidity, obtained in the reaction $^{124}$Sn + $^{124}$Sn at 50 MeV/u.
Results are shown for ImQMD calculations at b = 6 fm (thick line) and b = 8 fm (thin line)
and for SMF calculations at b = 6 fm (full circles) and b = 8 fm (open circles). 
A soft interaction is considered for the symmetry energy.
Right panel: N/Z of IMF's as a function of the reduced rapidity. Lines and symbols are
like in the left panel. 
Results corresponding to a stiff Asy-EOS are also shown for ImQMD (dashed line) and
SMF (crosses), for b=6 fm \cite{Zhang}.}
\end{figure}  

\subsection{Comparison with the predictions of different transport codes}
A detailed investigation of isospin equilibration has been recently undertaken
within transport codes based on the molecular dynamics (QMD) approach \cite{yingxun}.
In comparison to the transport model considered before, mainly 
describing mean-field mechanisms (the SMF model, 
see Eq.(1)),
such approaches, where nucleons are represented as individual wave packets of fixed
compact shape (usually taken as gaussians),
are well suited to describe  fluctuations and correlations, especially in
the exit channel of multifragmentation events, whereas some aspects of  
the mean-field dynamics may not be well accounted for.  
As shown in Ref.\cite{yingxun}, where charge equilibration is investigated for $Sn + Sn$ reactions
at 35 and 50 $MeV/u$, 
the ImQMD code predicts a quite different behavior with respect to SMF:
the isospin transport ratio is rather flat as a function
of the impact parameter.  
This seems to indicate that, even in the case of central collisions, the contact time
between the two reaction partners remains rather short, the dissipation mechanisms being mostly due
to many-body correlations and particle emission, rather than to mean-field effects. Thus the more explosive
dynamics could lead to the lower degree of isospin equilibration observed.
It should be noticed that the stronger impact of many-body correlations on the fragmentation
path in molecular dynamics approaches has also been evidenced, in the context of central
collisions, performimg a comparison between  the predictions of the Antisymmetrized Molecular Dynamics (AMD) model 
and of the SMF approach \cite{Akira}.

To examine more in detail the origin of the observed discrepancies, 
results concerning IMF ($Z>2$) properties, obtained with the SMF and ImQMD codes, are compared in figure 9.
In the left panel, the average total charge per event, associated with IMF's, is plotted
as a function of the reduced rapidity, for the reaction $^{124}$Sn + $^{124}$Sn at 50 MeV/u and 
impact parameters b = 6 and 8 fm. 
From this comparison it is clear that in ImQMD a larger number of light IMF's, distributed
over all rapidity range between PLF and TLF, are produced.
On the other hand, mostly binary or ternary events are observed in SMF, with light IMF's
located very close to mid-rapidity. Then the different reaction dynamics predicted by the two codes
may explain the different results seen for isospin equilibration especially
in semi-peripheral and central reactions (b $\approx$ 4-6 fm).
The fast ImQMD fragmentation dynamics inhibits nucleon exchange and charge equilibration,
though the energy loss is rather large (due to cluster emission).
On the other hand, in SMF dissipation
is dominated by mean-field mechanisms, acting over longer time intervals and leading to stronger
equilibration effects.

Results on the neutron content of the neck region are illustrated in the right panel of figure 9,
that shows the global N/Z of IMF's as a function of the reduced rapidity. As discussed above,
SMF calculations clearly predict a larger N/Z for IMF's produced at mid-rapidity,  
with respect to PLF and TLF regions (isospin migration effect). 
The effect is particularly pronounced in the case of the asystiff parametrization. 
On the contrary, ImQMD calculations predict a minimum of the N/Z ratio at mid-rapidity.
The reasons of these difference  need to be further investigated. 
Moreover, it would be interesting to perform detailed comparisons with
experimental data, also in consideration of the neck fragmentation analyses
recently appeared in the literature \cite{defilposter}.


\section{Asy-EOS at supra-saturation density: collective flows} 

Reactions with neutron-rich systems at intermediate energies (100-500 $AMeV$  ) are
of interest in order to have high momentum particles and to test regions
of high baryon and isovector density during the
reaction dynamics.
In such a context, it is important to consider 
momentum dependent effective interactions, which
essentially lead to the concept of effective masses. If also the isovector
component of the interaction is momentum dependent, one observes different
effective masses (i.e. effective mass splitting) for neutrons and protons.  
The problem of the precise determination of the Momentum Dependence in the Isovector
channel ($Iso-MD$) of the nuclear interaction 
is still very controversial and it would be extremely
important to get more definite experimental information \cite{BaoNPA735,rizzoPRC72},
looking at observables which may also be sensitive to the mass splitting.  

Transport codes are usually implemented with 
different $(n,p)$ momentum dependent interactions, see 
for instance \cite{BaoNPA735,rizzoPRC72}. 
This  allows one to follow the dynamical
effect of opposite n/p effective mass ($m^*$) splitting while keeping the
same density dependence of the symmetry energy \cite{isotr07}.

Let us consider semicentral ($b/b_{max}=0.5$) collisions of
 $^{197}$Au+$^{197}$Au at $400~AMeV$ \cite{vale08}.
In the interacting zone baryon densities about
$1.7-1.8 \rho_0$ can be reached in a transient time of the order of 15-20 fm/c. 
The system 
is quickly expanding and the freeze-out time is around 50 fm/c.
A rather abundant particle emission is observed over this time scale. 
Here it is very interesting to study again the collective response of the system.
Collective flows are very good candidates since they are 
expected to be 
rather sensitive to the momentum
dependence of the mean field, see \cite{DanielNPA673,baranPR}.
The transverse flow, 
$V_1(y,p_t)=\langle \frac{p_x}{p_t} \rangle$,
where $p_t=\sqrt{p_x^2+p_y^2}$ is the transverse momentum and $y$
the rapidity along the beam direction, 
provides information on the anisotropy of 
nucleon emission on the reaction plane.
Very important for the reaction dynamics is also the elliptic
flow,
$V_2(y,p_t)=\langle \frac{p_x^2-p_y^2}{p_t^2} \rangle$.
 The sign of $V_2$ indicates the azimuthal anisotropy of the emission:
on the reaction
plane ($V_2>0$) or out-of-plane ($squeeze-out,~V_2<0$)
\cite{DanielNPA673}.


\begin{figure}[h]
\resizebox{0.5\textwidth}{!}{%
\includegraphics{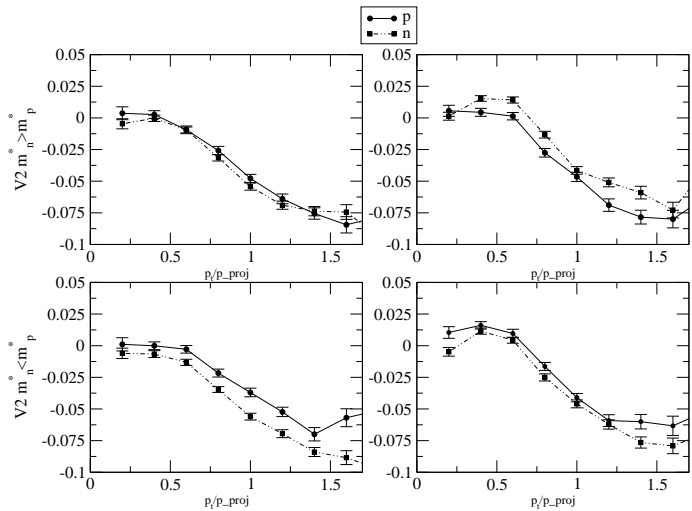}} 
\caption{\label{v2ypt400}
Proton (thick) and neutron (thin) $V_2$ flows in a 
semi-central
reaction Au+Au at 400 $AMeV$.
Transverse momentum dependence at midrapidity, $\mid y_0 \mid <0.3$.
Upper curves for $m_n^*>m_p^*$, lower curves for
the opposite splitting $m_n^*<m_p^*$. Left: asystiff. Right: 
asysoft \cite{vale08}.}
\end{figure}  
In 
figure \ref{v2ypt400}, 
we plot the elliptic flow of emitted neutrons and protons, for different asy-stiffness
and effective mass splitting choices. 
We are now exploring density regions above normal density. Therefore we expect a larger
neutron repulsion in the asystiff case, corresponding to the larger symmetry energy value, see 
figure 1. 
Indeed in figure 10 we observe a larger (negative) squeeze-out for neutrons in the asystiff case
(compare left and right panels). 
Moreover, the  $m^*_n < m^*_p$ case will favor the neutron repulsion,
leading to a larger squeeze-out for neutrons, compare top and bottom panels.  
In particular, in the asysoft case (on the right) we see an inversion of the 
neutron/proton squeeze-out
at mid-rapidity for the two effective mass-splittings. 

Actually, we observe a rather interesting interplay between the effects
linked to the symmetry energy and to the mass splitting: 
a larger (smaller) neutron effective mass may compensate the larger (smaller) neutron repulsion
corresponding to the asystiff (asysoft) case.  
Indeed the $m^*_n < m^*_p$ case,
with the soft Asy-EOS, yields very similar results of the $m^*_p < m^*_n$ case with the 
stiff Asy-EOS.

It seems to be difficult to conclude on the properties of the effective
interaction (asystiffness and MD) just from the analysis of one single obervable. 
However, coupling the flow information to the study of 
other observables,   it would be possible to reach more definitive constraints
of the effective interaction. For instance, in the considered beam energy range, 
the N/Z content of the particle emission  looks 
particularly sensitive just to the sign of mass splitting, rather than to the asy-stiffness
\cite{vale08}. Hence it would be very interesting the combine the information coming from particle flows
and yields. 
Recent experimental analyses look very promising in this direction  \cite{Paolo,Coz}.
Due to the difficulties in
measuring neutrons, one could also investigate the difference between
light isobar (like $^3$H vs. $^3$He) flows and yields. 
We still expect to see effective mass splitting effects \cite{vale08}. 


\section{Hadron-Quark transition at high isospin and baryon density}

In heavy ion collisions at beam energies in the AGeV range, rather high
density regions can be reached, opening the possibility that new
degrees of freedom come into play. 
This kind of collisions is usually described within relativistic mean-field
(RMF) models and transport theories \cite{baranPR}.
In neutron-rich systems, the transition from the nuclear (hadron) 
to the quark deconfined   
 (quark-gluon plasma) phase could take place even at the density
and temperature conditions reached along the collision dynamics. 
This kind of transition is also of large interest in the study of 
neutron stars. 


At high temperature $T$ and small
quark chemical potential $\mu_q$ 
lattice-QCD calculations provide a
valuable tool to describe such transition. 
The transition
appears of continuous type (crossover) with a critical temperature
$T_c$ around 170-180 MeV. Isospin and other properties of the hadron
interaction appear not relevant here.

However the fundamental lattice calculations suffer serious problems
at large chemical potentials and the validity of the results at
$\mu_q/T_c > 1$ is largely uncertain ~\cite{Fukushima11}. Some
phenomenological effective models have been introduced, like the
MIT-Bag ~\cite{Chodos74} and the more sophisticated Nambu-Jona
Lasinio (NJL)~\cite{Nambu61,Hatsuda94,Buballa05} and Polyakov-NJL
~\cite{Fukushima04,Ratti06,Schaefer10} models, where the chiral and
deconfinement dynamics is accounted for. We remark here that only
scalar interactions are generally considered in the quark sector. The
transition at low-$\mu$ is well in agreement with l-QCD results
however still properties of the hadron sector are not included and
so the expected transition at high baryon and isospin density cannot
be trusted.

In order to overcome the problem and to get some predictions about
the effect of the transition in compact stars
~\cite{Glendenning92,Glendenning98,Burgio02,Shao10,Shao110,Dexheimer101,Dexheimer102} and high energy
heavy ion collisions ~\cite{Muller97,Toro06,Torohq11,Liu11,Cavagnoli10,Shao111,Shao112}, recently Two-Phase
(Two-Equation of State) models have been introduced where both
hadron and quark degrees of freedom are considered, with particular
attention to the transition in isospin asymmetric matter.

\begin{figure}[htbp]
\resizebox{0.5\textwidth}{!}{%
  \includegraphics{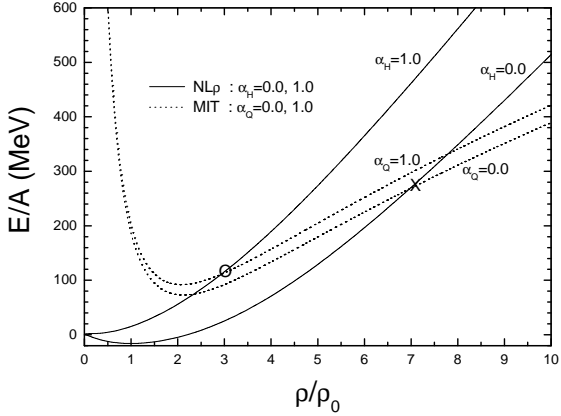}
}
\caption{\label{T0isotrans}{(Color online) Energy per particle at
zero temperature for the nucleon (solid) and quark (dotted) sectors
for the cases of isospin symmetric ($\alpha=0$) and "pure neutron"
($\alpha=1$ matter) ~\cite{Torohq11}.}}
\end{figure}

A simple argument can be used for the expected isospin effects. In
figure \ref{T0isotrans} we report the energy per particle at $T=0$ vs.
the baryon density for the hadron (RMF EoS, \cite{baranPR}) and the
quark (MIT-Bag) matter in the isospin symmetric ($\alpha_{H,Q}=0$)
and "pure neutron" ($\alpha_{H,Q}=1$) cases. For energetic reasons
we roughly expect the transition to appear around the crossings
\cite{Torohq11}. We see large isospin effects on the
transition region: for symmetric matter we are around $7$ times the
saturation density, for the completely isospin asymmetric matter we
move down to $3 \rho_0$. This is very interesting for neutron stars,
as well as for heavy ion collisions in the $ AGeV$ energy range. We
note that this effect is due to the lower symmetry pressure in the
quark phase, where the symmetry energy comes only from the kinetic
Fermi contributions due to the lack of isovector
interactions.

When a mixed (coexistence) phase of quarks and hadrons is considered,
the Gibbs conditions (thermal, chemical and mechanical equilibrium)
\begin{eqnarray}\label{tcm}
& &\mu_B^H(\rho_B^{},\rho_3^{},T)=\mu_B^Q(\rho_B^{},\rho_3^{},T)\nonumber\\
& &\mu_3^H(\rho_B^{},\rho_3^{},T)=\mu_3^Q(\rho_B^{},\rho_3^{},T)\nonumber\\
& &P^H(\rho_B^{},\rho_3^{},T)=P^Q(\rho_B^{},\rho_3^{},T),
\end{eqnarray}
should be fulfilled ~\cite{Glendenning92}. In Eqs.~(\ref{tcm}),
$\rho_B^{}=(1-\chi)\rho_B^{H}+\chi \rho_B^{Q}$ is the mean baryon
density and  $\rho_3^{}=(1-\chi)\rho_3^{H}+\chi \rho_3^{Q}$ is the
isospin density, where $\chi$ is the quark fraction.
$\rho_B^{H,Q}/\rho_3^{H,Q}, \mu_B^{H,Q}/\mu_3^{H,Q}$ are
baryon/isospin densities and corresponding chemical potentials in
the two phases. $P^{H,Q}$ indicates the pressure in the two phases.  

In heavy-ion collisions, for a given isospin asymmetry of the considered
experiment, the global asymmetry parameter $\alpha$ 
%
\begin{eqnarray}
& &   \alpha\equiv-\frac{\rho_{3}^{}}{\rho_{B}^{}}= - \frac{(1-\chi)\rho_3^{H}+
\chi \rho_3^{Q}}{(1-\chi)\rho_B^{H}+\chi \rho_B^{Q}},
\end{eqnarray}
keeps constant  according to the isospin charge conservation in the strong interaction,
but the local asymmetry parameters $\alpha^H,\alpha^Q$ in the
separate phases can vary with $\chi$, which determines the energetically stable state of the system.
For details, one can
refer to Refs.~\cite{Torohq11,Shao111,Shao112}.

\begin{figure}[htbp]
\resizebox{0.4\textwidth}{!}{%
  \includegraphics{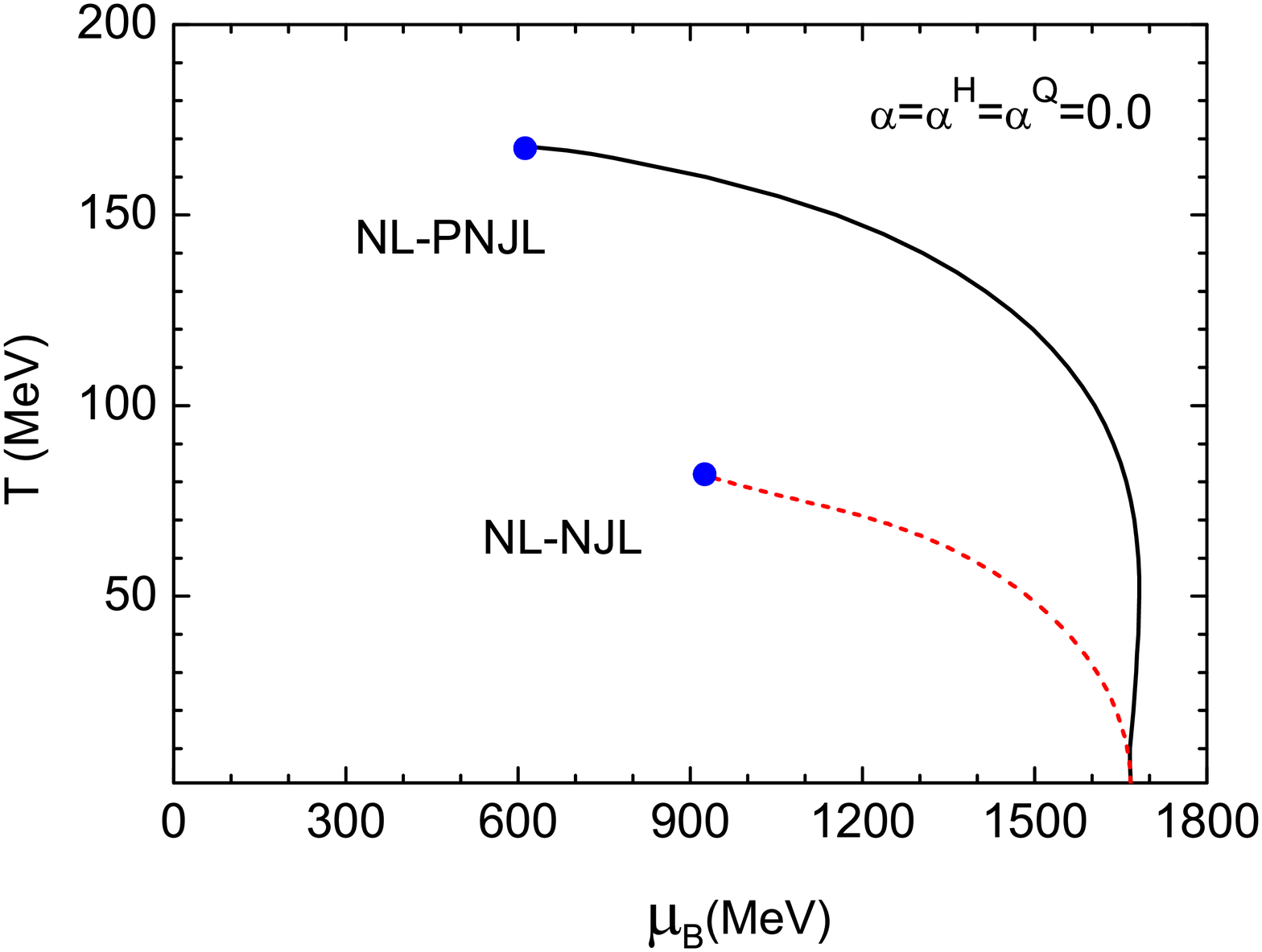}
}
\caption{\label{fig:T-Mu-with-delta-alpha=0} (Color on line) Phase
diagram in $T-\mu_B^{}$ plane for symmetric matter \cite{Shao112}.}
\end{figure}

\begin{figure}[htbp]
\resizebox{0.4\textwidth}{!}{%
  \includegraphics{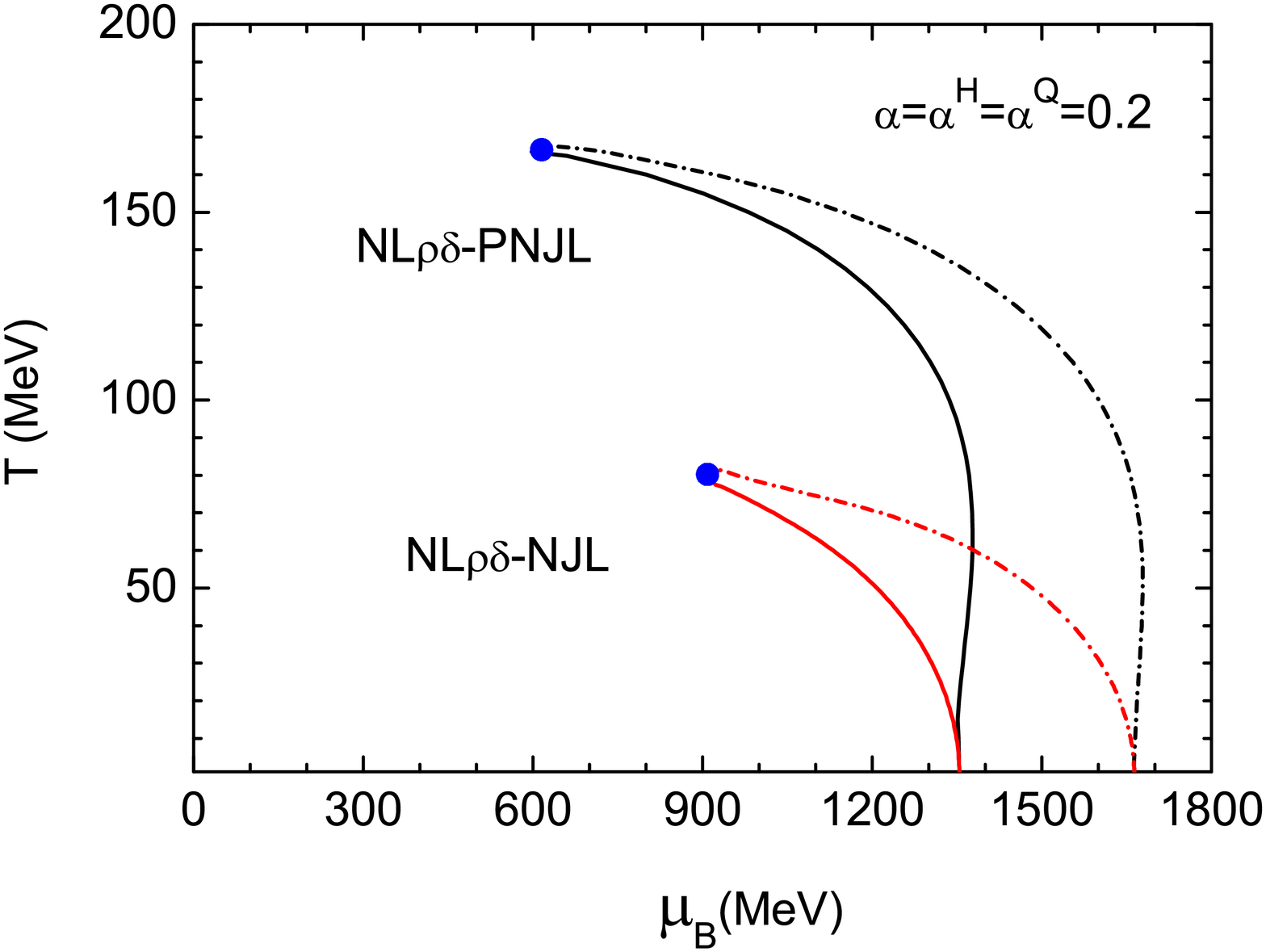}
}
\caption{\label{fig:T-Mu-with-delta-alpha=02} (Color on line) Phase
diagram in  $T-\mu_B^{}$ plane for asymmetry matter with the global
asymmetry parameter $\alpha=0.2$ \cite{Shao112}.}
\end{figure}

In figures ~\ref{fig:T-Mu-with-delta-alpha=0} and
~\ref{fig:T-Mu-with-delta-alpha=02} we plot the phase transition
curves with the Hadron-NJL and Hadron-PNJL models. At low
temperatures a clear earlier onset of the transition is observed for
isospin asymmetric matter (see full lines of figure~\ref{fig:T-Mu-with-delta-alpha=02}).
 For the NJL model
with only chiral dynamics, no physical solution exists when the
temperature is higher than
 $\sim80$
MeV.  The corresponding temperature is enhanced to about $\sim166$
MeV with the Hadron-PNJL model,
 which is closer to the phase transition (crossover)
temperature given by full lattice calculations at zero or small
chemical potential.
 A general observation is that the coupling to the Polyakov loop
 field is essentially reducing the quark/antiquark distribution functions
 due to the confinement constraint. As a consequence the quark pressure at
 high temperature will be also reduced and this will make possible
 a hadron-quark first order transition at higher temperatures.

From figure~\ref{fig:T-Mu-with-delta-alpha=02} we remark that in both
cases the region around the $Critical-End-Points$ is almost not
affected by isospin asymmetry contributions, which are relevant at
lower temperatures and larger chemical potentials.


\begin{figure}[htbp]
\resizebox{0.4\textwidth}{!}{%
  \includegraphics{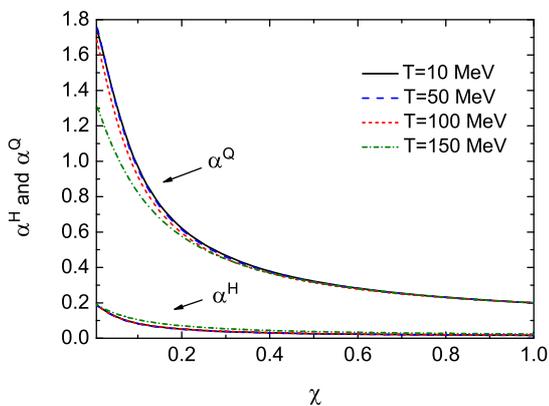}
}
\caption{\label{fig:kai-alphaQ-NLrho-with-delta-alpha=02} (Color on
line)
 The behavior of
local asymmetric parameters $\alpha^H$ and $\alpha^Q$ in the mixed
phase for several values of temperature. Parameter set
$NL\rho\delta$ is used in the calculation \cite{Shao112}.}
\end{figure}

For symmetric matter there is only one phase-transition line in the
$T-\mu_B^{}$ plane, i.e. the phase transition curve is independent
of the quark fraction $\chi$. At variance, for asymmetric matter,
the phase transition curve varies for different quark fraction
$\chi$. The phase transition curves  in
figure~\ref{fig:T-Mu-with-delta-alpha=02} are obtained with $\chi=0$
and $1$,  representing the beginning and the end of the hadron-quark
transition, respectively. The reason is that we have an
important Isospin Fractionation (Distillation) effect, i.e., an
enhancement of the isospin asymmetry in the quark component inside
the mixed phase, as reported in
figure~\ref{fig:kai-alphaQ-NLrho-with-delta-alpha=02}, where the
asymmetry parameters in the two components are plotted vs. the quark
fraction $\chi$.
In asymmetric matter at a fixed temperature, along the transition path, i.e. increasing the
quark fraction, pressure and chemical potential change and the two
coexisting phases have different asymmetry.  

%

These features of the local asymmetry may lead to some observable
effects in the hadronization during the expansion phase of heavy ion
collisions, such as an inversion in the trend of emission of neutron
rich clusters, an enhancement of $\pi^-/\pi^+$, $K^0/K^+$ yield
ratios in high-density regions, as well as an enhancement of the
production of isospin-rich resonances and subsequent decays, for
more details see Refs.~\cite{Torohq11,Shao111,Shao112}. Such signals
are possible to be probed in the newly planned facilities, such as
FAIR at GSI-Darmstadt and NICA at JINR-Dubna.

\subsection{Isoscalar vector quark interaction and existence of hybrid neutron stars}

Inside this frame it appears natural to study the role of vector
interactions in the quark effective models \cite{Shao12,Shao13}. We
remind that in nuclear matter the vector interactions lead to
fundamental properties, like the saturation point and the symmetry
energy in isospin asymmetric sistems. We discuss now the results
obtained when the isoscalar--vector interaction channel in the quark
sector is turned on in the (P)NJL models.
 With increasing the ratio $R_V=G_\omega/G$ of the vector/scalar coupling constants,
 due to the repulsive contribution of the
isoscalar--vector channel to the quark energy and, as a consequence,
to the chemical potential, the phase-transition curves are moving
towards higher values of density/chemical potential \cite{Shao13}.
The larger repulsion in the quark phase is essential for the
existence of massive hybrid neutron stars. The limit appears to be
the impossibility of reaching the onset densities of the mixed phase
in the inner core for large values of the vector coupling.
\begin{figure}[htbp]
\resizebox{0.4\textwidth}{!}{%
  \includegraphics{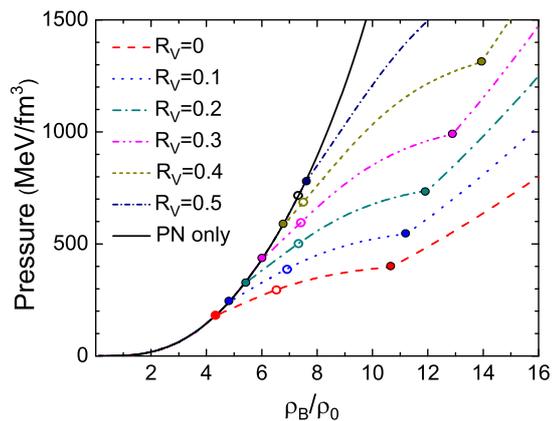}
}
\caption{\label{fig:EoS1}(color on line) EoSs of neutron star matter
without and with a hadron-quark phase transition for different
isoscalar-vector interaction coupling $G_V$. For each value of
$R_V$, the two solid
 dots with the same color indicate the range of the mixed phase, and the cycle marks
  the largest pressure that can be reached in the core of the neutron star
\cite{Shao13}.}
\end{figure}

We present in figure~\ref{fig:EoS1} the EoS of neutron star matter
without (PN only) and with the hadron-quark phase transition for different
strength of the isoscalar-vector interaction. For each value of $R_V
= G_V/G$, the two solid dots with the same color indicate the range
of the mixed phase, and the cycle marks the largest pressure
that can be reached in the core of neutron star by solving the 
Tolman-Oppenheimer-Volkoff (TOV)
equation.  With increasing the vector strength in the quark sector the
onset of the transition is moving to higher densities since the
quark pressure is also increasing. A massive hybrid neutron star can
be supported in the range 0.1-0.3 of the $R_V$ ratio and good
agreement with recent data for the Mass-Radius relation are also
obtained \cite{Shao13}.



\section{Conclusions}
We have reviewed some aspects of the rich phenomenology associated with  
heavy ion reactions, from which interesting hints
are emerging to constrain the nuclear EOS and, in particular, the largely 
debated density behavior of the symmetry energy.
 
Information on the low density region 
can be accessed in reactions from low to Fermi energies, 
where collective excitations and fragmentation mechanisms are dominant.
Results on isospin sensitive observables have been presented.
In particular, we have concentrated our analysis on the charge equilibration
mechanism (and its relation to energy dissipation) and on the neutron-enrichment of the neck region 
in semi-peripheral reactions.  From the study of the latter mechanisms,
for which new experimental evidences have recently appeared \cite{defilposter,Emmanuelle}, hints are
emerging towards a moderately stiff behavior of the symmetry energy around normal 
density (L$\approx$75 MeV).       
This is compatible with recent results from structure data,
see for instance the review article \cite{structure_data}.
We have also tried to address the problem of the model dependence of 
the results, 
suggesting possible ways to better ascertain, through the 
comparison with all available experimental observables, 
the overall reaction dynamics, 
thus increasing the robustness of the extracted symmetry energy information 
 \cite{Akira,Zach,coup,paolo_new}.

The greatest theoretical uncertainties concerns the high density 
domain, 
that has the largest impact on the
understanding of the properties of neutron stars. 
In particular, a large symmetry pressure would favor the onset
of a quark phase in the inner core of neutron stars \cite{Shao13}.
This regime can be explored in terrestrial laboratories by using 
intermediate energy and relativistic
heavy ion collisions of charge asymmetric nuclei. 
Collective flows,
cluster and meson production are promising observables.  
Moreover, if the mixed phase can be reached, 
        at high baryon density, in collisions of isospin asymmetric 
        heavy ions, 
an isospin fractionation effect, 
leading to a more asymmetric 
        quark component, is expected to appear. 
Observable effects will be present in the subsequent 
        hadronization.

A considerable amount of work has already been done in the symmetry energy domain.  
In the near future, thanks to the availability of both stable and rare
isotope beams, more global analyses, also based on new sensitive 
observables, together with the comparison and the improvement of the available 
theoretical models, 
are expected to provide further stringent constraints. 
\noindent
{\bf Acknowledgements}

\noindent 
We warmly thank G.Y. Shao for the fruitful collaboration, during his stay at
LNS-Catania, on hadronic/quak matter phase transition and neutron star 
properties. 

This work for V. Baran was supported by a grant of the Romanian National
Authority for Scientific Research, CNCS - UEFISCDI, project number PN-II-ID-PCE-2011-3-0972.




\end{document}